\documentclass[prb,twocolumn,10pt,aps]{revtex4-2}

\usepackage[utf8]{inputenc}
\usepackage[T1]{fontenc}
\usepackage{lmodern}

\usepackage{amsmath}
\usepackage{amsfonts}
\usepackage{amssymb}
\usepackage{mathtools}
\usepackage{physics}
\usepackage{bbold}

\usepackage{hyperref}
\hypersetup{
	colorlinks   = true, %Colours links instead of ugly boxes
	urlcolor     = blue, %Colour for external hyperlinks
	linkcolor    = red, %Colour of internal links
	citecolor   = blue %Colour of citations
}

\usepackage{xcolor}

\newcommand{\half}[1][1]{\frac{#1}{2}}
\newcommand{\inv}[1]{\frac{1}{#1}}

\newcommand{\ir}{\mathrm{i}}
\newcommand{\er}{\mathrm{e}}
\newcommand{\del}{\partial}

\newcommand{\vk}{\vb{k}}

\newcommand{\thetak}{\theta_{\vk}}
\newcommand{\phik}{\varphi_{\vk}}
\newcommand{\xik}{\xi_{\vk}}
\newcommand{\zetak}{\zeta_{\vk}}

\newcommand{\PT}{$\mathcal{PT}$}

\newcommand{\Vv}{\mathcal{V}}

\newcommand{\vphi}{\vb*{\phi}}
\newcommand{\vPi}{\vb*{\Pi}}
\newcommand{\rc}[1]{r_{{#1}\mathrm{c}}}
\newcommand{\reff}{r^{\mathrm{eff}}}

\usepackage[normalem]{ulem}

\begin{document}

\title{Universal scaling at a prethermal dark state}
\author{Marvin Syed}
\email{ms2808@cam.ac.uk}
\affiliation{Department of Applied Mathematics and Theoretical Physics, University of Cambridge, Cambridge CB3 0WA, United Kingdom}
\affiliation{Institut f\"ur Theoretische Physik, Universit\"at Heidelberg, 69120 Heidelberg, Germany}
\author{Tilman Enss}
\affiliation{Institut f\"ur Theoretische Physik, Universit\"at Heidelberg, 69120 Heidelberg, Germany}
\author{Nicol\`o Defenu}
%\email{ndefenu@phys.ethz.ch}
\affiliation{Institute for Theoretical Physics, ETH Z\"urich, Wolfgang-Pauli-Str. 27, 8093 Z\"urich, Switzerland}
\begin{abstract}
Recent experimental and theoretical progress as well as the prospect of commercially viable quantum technologies have inspired great interest in the study of open quantum systems and their dynamics. Many open quantum systems are well described by an effective non-Hermitian Hamiltonian generating a time evolution that allows eigenstates to decay and dissipate to the environment. In this framework, quantum coherent scaling is traditionally tied to the appearance of dark states, where the effect of dissipation becomes negligible. Here, we discuss the universal dynamical scaling after a sudden quench of the non-Hermitian $O(N)$ model Hamiltonian. While universality is generally spoiled by non-Hermiticity, we find that for a given set of internal parameters short-time scaling behaviour is restored with an initial slip exponent profoundly different from that of closed quantum systems. This result is tied to the compensation of dissipation by interaction effects at short times leading to a prethermal dark state, where coherent many-body dynamics can be still observed.
\end{abstract}
\maketitle

\section{Introduction} The continuous advancement in the realisation and study of quantum systems has recently stimulated an enormous interest in the development of quantum technologies, which exploit  entanglement and coherence properties to reach a quantum advantage over analogous classical devices\,\cite{zoller_quantum_1995, verstraete_quantum_2009, blatt_quantum_2012, degen_quantum_2017}. In order to achieve such goals, it is often necessary to realise large-scale tunable systems, whose parameters can be modified by suitable manipulation of the external environment and which retain quantum coherence also at the many-body level. Therefore, the number of experimental applications in the field of driven-dissipative many-body systems is steadily increasing\,\cite{barreiro_open_2011, krauter_entanglement_2011, yan_observation_2013,
carr_nonequilibrium_2013, carusotto_quantum_2013, mivehvar_cavity_2021,monroe_programmable_2021,defenu_longrange_2021}.

Given the many-body nature of these realisations, several questions have arisen regarding collective phenomena and (dynamical) critical behaviour in open systems\,\cite{wouters_absence_2006, mitra_nonequilibrium_2006, deng_exciton_2010, torre_quantum_2010, houck_chip_2012, sieberer_dynamical_2013, carusotto_quantum_2013, nagy_nonequilibrium_2015}. While the presence of dissipation typically reduces this bulk critical behaviour to the classical equilibrium universality\,\cite{sieberer_dynamical_2013, carusotto_quantum_2013, nagy_nonequilibrium_2015}, genuine out-of-equilibrium quantum scaling can be observed  for systems trapped into dark states\,\cite{marino_driven_2016}. 
Universal dynamics is also attained by a many-body system during its approach to thermal equilibrium in the vicinity of a dynamical critical point. There, the dynamics often enters a \emph{prethermal} quasi-stationary state before finally approaching the thermal state\,\cite{mori_thermalization_2018}. During the prethermal stage the system exhibits universal scaling laws\,\cite{gambassi_quantum_2011, sciolla_quantum_2013, chandran_equilibration_2013, maraga_aging_2015, chiocchetta_short-time_2015, chiocchetta_dynamical_2017} which are reminiscent of ageing in classical dissipative systems\,\cite{janssen_new_1989,calabrese_ageing_2005}.

In the case of driven-dissipative systems, prethermal universality appears in analogy with the case of closed quantum systems\,\cite{maraga_aging_2015}, but with peculiar scaling exponents that have no counterpart in the isolated case, as shown in the following.
To our knowledge such novel prethermal scaling manifests itself only for specific values of the internal system parameters, where dynamical quantum correlations compensate the effect of dissipation, at least at short times, leading to an effective prethermal dark state.

Here, this phenomenon is analysed in the prototypical example of dissipative $O(N)$ field theories. These are described in terms of non-Hermitian Hamiltonians, in analogy with other cases where a finite coupling to the environment yields effective non-Hermitian Hamiltonians\,\cite{dalibard_wave_1992, moiseyev_non_2011,brody_mixed_2012,daley_quantum_2014,lee_entanglement_2014,wu_observation_2019,ashida_non_2020}.
Remarkably, some non-Hermitian Hamiltonians can still possess a real energy spectrum, namely, when the Hamiltonian and its eigenvectors are symmetric under parity and time-reversal, also known as \PT\ symmetry\,\cite{bender_real_1998, bender_complex_2002, bender_making_2007}. Often these systems possess a so-called exceptional point where the energy spectrum switches over from real to complex eigenvalues. This is often referred to as spontaneous \PT\ symmetry breaking. Various \PT-symmetric systems have been studied\,\cite{froml_fluctuation-induced_2019, graefe_mean-field_2008, begun_phase_2021, hu_probability-preserving_2012, para_probing_2021, fring_solvable_2018, frith_time-dependence_2020, wang_spontaneous_2015, dora_kibble-zurek_2019, lin_wide-range-tunable_2017} and experimentally realised in recent years\,\cite{klaiman_visalization_2008, guo_observation_2009, ruter_observation_2010, bittner_PT_2012, hang_symmetry_2013, fleury_invisible_2015, peng_anti-paritytime_2016, schindler_experimental_2011, zhang_observation_2016, bender_observation_2013}.
	
 More recently, theoretical studies have pushed the application of non-Hermitian Hamiltonians beyond the framework of few-body systems, leading to pioneering studies on quantum criticality in non-Hermitian many-body systems\,\cite{lee_heralded_2014,yamamoto_theory_2019,ashida_non_2020}. Interestingly, in most of these studies the breaking of Hermitian dynamics is not related to the explicit presence of the environment, but rather to performing repeated projective measurements, whose relation to dark states recently raised a widespread interest\,\cite{ashida_non_2020,thiel_dark_2020,liu_driving_2021}.Yet,  the possibility to achieve universal scaling regimes out of equilibrium has not been investigated so far and several questions regarding dynamical criticality of non-Hermitian systems remain open.

As was shown in Refs.\,\cite{fring_solvable_2018,frith_time-dependence_2020}, \PT-symmetric Hamiltonians are more naturally studied in an explicitly time-dependent setting. In this work we show that the contribution of non-Hermitian dynamics to prethermal universal scaling leads to a rich phenomenology, which has not been observed before. The possibility to observe prethermal dark states and their corresponding universal scaling in $O(N)$ vector models underlines the universal nature of our findings: indeed, $O(N)$ field theories are well-suited to describe many different physical systems in particle physics and cosmology as well as in condensed matter \cite{sachdev_2011}, and prethermal scaling has been discussed for the universality class of the Ising ($N=1$) and Heisenberg ($N=3$) models\,\cite{janssen_from_1992}. In the limit $N\to\infty$ the $O(N)$ model becomes effectively quadratic\,\cite{moshe_quantum_2003}, while still retaining important qualitative features of the equilibrium phase diagram for finite $N$\,\cite{berges_controlled_2002, chandran_equilibration_2013, babadi_far_2015, weidinger_dynamical_2017,lerose_chaotic_2018, lerose_prethermal_2019}. Large-$N$ approaches have been shown to coherently describe dynamical phase transitions\,\cite{weidinger_dynamical_2017,syed_dynamical_2021} and dynamical scaling\,\cite{maraga_aging_2015, chiocchetta_short-time_2015, chiocchetta_dynamical_2017} in general critical systems.  

\section{Model and Dyson map} The Hamiltonian for our model consists of three parts,
\begin{align}\label{totham}
\mathcal{H}(\lambda,\{r_{1},r_{2}\},\{u_{1},u_{2}\})= H_1  + H_2  + H_{12}(\lambda) 
\end{align}
where
\begin{align}
\label{ham_eq}
H_i  =  \int_x \qty[\half\vb*{\Pi}_{i}^2 + \half\qty(\grad{\!\vb*{\phi}_{i}})^2 + \half[r_i] \vb*{\phi}_{i}^2 +\frac{u_i}{4!N} \qty(\vb*{\phi}_{i}^2)^2]
\end{align}
is the usual $O(N)$ model Hamiltonian for two $N$-component scalar fields
$\vb*{\phi}_1 = (\phi_1^1, \phi_1^2, \dots, \phi_1^N)$ and $\vb*{\phi}_2 = (\phi_2^1, \phi_2^2, \dots, \phi_2^N)$ with $\vb*{\Pi}_i$ their respective conjugate momenta obeying $\comm{\phi_i^a(\vb{x},t)}{\Pi_j^b(\vb{y},t)} = \ir \delta(\vb{x}-\vb{y})\delta_{ij}\delta_{ab}$. The symbol $\int_x = \int \dd^d x$ indicates integration over space. The parameters $r_i$ control the bare mass of the system while the $u_i$ control the strength of the quartic self-interactions.  The final term
\begin{align}
\label{non_herm_term}
H_{12}(\lambda) = \ir\lambda\int_x \qty[\vb*{\phi}_1 \vdot \vb*{\phi}_2]
\end{align}
is an imaginary coupling between the two fields controlled by the real parameter $\lambda$. The total Hamiltonian $H$ is invariant under the antilinear transformation $\mathcal{PT}:\vphi_1 \to \vphi_1, \vphi_2 \to -\vphi_2, \vPi_1 \to -\vPi_1, \vPi_2 \to \vPi_2, \ir \to -\ir$.

A suitable microscopic model that realises the Hamiltonian in Eq.\,\eqref{totham} can be achieved using continuously monitored ultracold atoms as in Ref.\,\cite{ashida_parity-time-symmetric_2017}.
There, excitations of the ground state to a short-lived excited state with near-resonant light create an effective imaginary mass. 
We propose a setup using atoms with multiple ground and excited states with appropriate transitions between them. 
Preliminary calculations show that the effective Hamiltonian describing this system has the same form of our Eq.\,\eqref{totham}, where the fields now describe the continuously monitored ground state atoms.  The virtual excitation and de-excitation of ground-state atoms together with the fast decay of excited state atoms realises the imaginary coupling in Eq.\,\eqref{non_herm_term}.

The system still exhibits a phase transition since $H_{12}$ preserves the $O(N)$ symmetry and spontaneous symmetry breaking can occur\,\cite{syed_equilibrium_inpreparation}. We note that this model bears some similarities to the one proposed in Ref.\,\cite{gagel_universal_2014}, but instead of an $O(N)$ field coupled to a bath of harmonic oscillators we have an $O(N)$ field with an imaginary coupling to another $O(N)$ field. Below we show that in our case, after taking the $N\to\infty$ limit, the second field effectively decouples from the first one and one is left with an effective theory for a single scalar field with an additional time dependence in the mass coefficient arising from the non-Hermitian $H_{12}$ term.

\begin{figure}
\includegraphics[width=\linewidth]{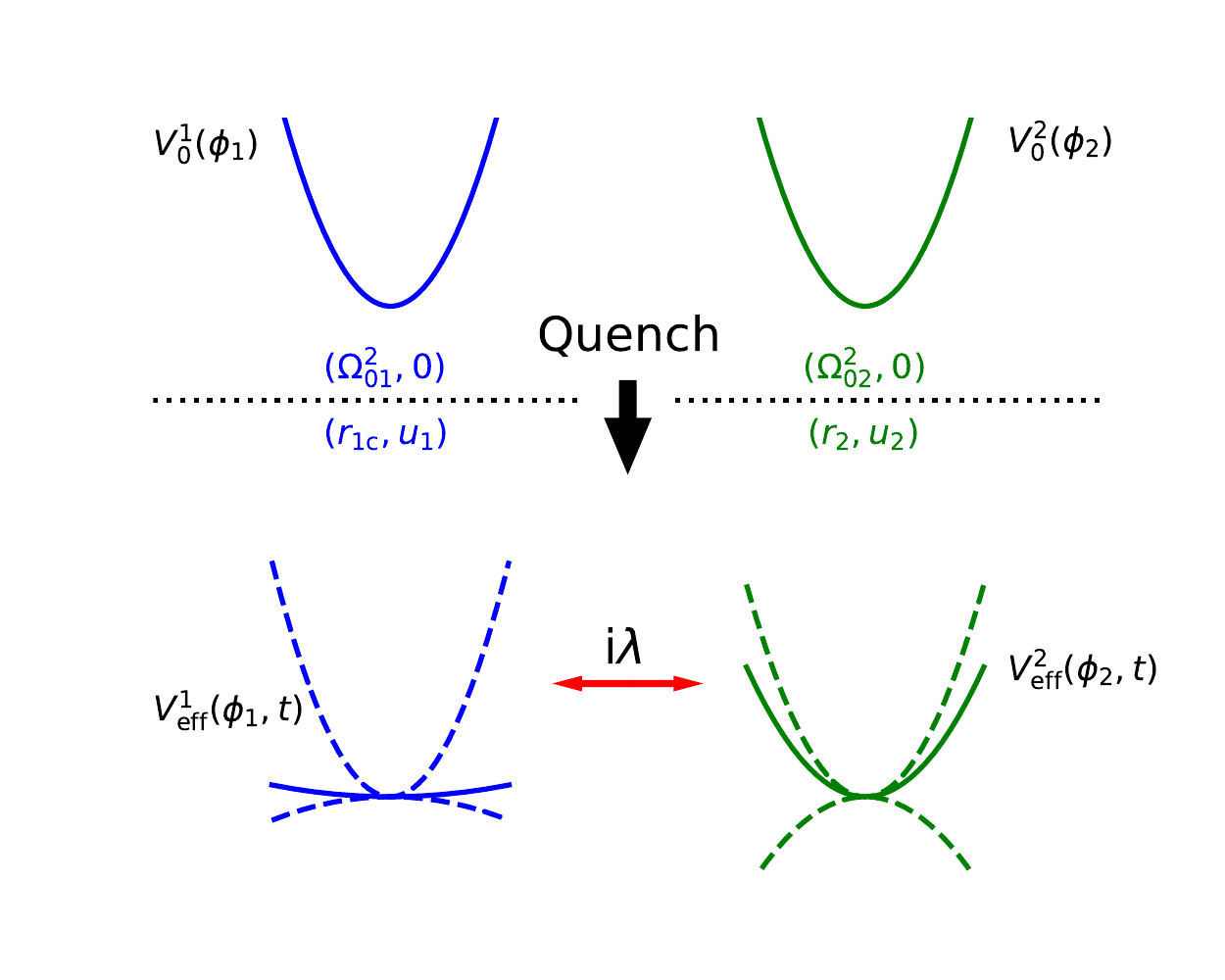}
\caption{Our model for a non-Hermitian system consists of two $O(N)$ fields which are initially prepared in the ground state of the free, non-interacting Hamiltonian with potential $V_0^i(\phi_i) = \Omega_{0i}^2 \phi_i^2 / 2$. At $t=0$ we switch on quartic self-interactions $u_1, u_2$, as well as the non-Hermitian coupling $\ir\lambda$ between the fields. The interactions lead to an effective quadratic potential $V_\mathrm{eff}^i(\phi_i,t) = \reff_i(t) \phi_i^2 / 2$ with time-dependent effective mass $\reff_i(t)$. Furthermore, we choose the post-quench mass parameter $\rc{1}$ such that the long-time limit of $\reff_1(t)$ vanishes.  In the lower panel, the solid line represents the change in the bare potential, while the dashed lines represent the oscillations of the effective potential $V_\mathrm{eff}^i(\phi_i,t)$ caused by the time-dependent effective mass.}
\label{fig:quench}
\end{figure}
For the quench, the system is prepared in the disordered (paramagnetic) ground state of the noninteracting initial Hamiltonian $\mathcal{H}(0,\qty{\Omega_{01}^2,\Omega_{02}^2}, \qty{0,0})$. At $t=0$ interactions as well as the imaginary coupling are turned on and the system evolves under the Hamiltonian $\mathcal{H}(\lambda, \qty{\rc{1}, r_2}, \qty{u_1, u_2})$. It was shown previously that the standard $O(N)$ model can undergo a dynamical phase transition (DPT) at a critical value $r=r_\mathrm{c}$\,\cite{maraga_aging_2015}, and following a similar analysis the same phenomenon is found to occur also in the non-Hermitian case\,(see Appendix).

The symmetry of the initial state implies that the expectation value of the two-point function $\ev{\phi_i^a \phi_j^b}$ does not develop mixed terms between different field components, i.e., $\ev{\phi_i^a \phi_j^b}=\delta_{ab}\ev{\phi_i^a \phi_j^a}$ for a single field component $\phi_i^a$.
From now on we will drop the component indices and treat $\phi_i$ as an arbitrary field component. The model in Eq.\,\eqref{ham_eq} is interacting and cannot be solved exactly. We then employ a conventional approximate solution where the quartic interaction is assumed to decouple into a self-consistent quadratic term, i.e.,
\begin{align}
\label{eq_dec}
(\vphi_i^2)^2 \to 2(N+2) \ev{(\phi_i)^2}\vphi_i^2 - N(N+2) \ev{(\phi_i)^2}^2.
\end{align}
We have $\langle\phi_{i}\rangle=0$ consistent with the system being in the symmetric phase, and the prefactors follow from Wick’s theorem\,\cite{wick1950evaluation}.

The decoupling in Eq.\,\eqref{eq_dec} coincides with the Hartree-Fock approximation of an interacting $O(N)$ field theory and can be rigorously justified in the $N\to\infty$ limit\,\cite{sotiriadis2010quantum}. This leads to the following effective Hamiltonian for each field component separately,
\begin{align}
\label{eff_ham}
\begin{split}
	H_\mathrm{eff}(t) = \int_x \Bigl\{ \half \sum_{i=1,2} 
		\bigl[ & \Pi_i^2 + (\grad{\phi_i})^2 \\ 
		       & + r_{i,\text{eff}}(t) \phi_i^2 \bigr] + 
		       \ir\lambda \phi_1 \phi_2
\Bigr\}
\end{split}
\end{align}
with $r_{i,\text{eff}}(t)$ fulfilling the self-consistency equation
\begin{align}\label{eq:sc}
r_{i,\text{eff}}(t) = r_i + \frac{u_i}{6} \ev{\phi_i^2(t)} \ .
\end{align}
Eqs.\,\eqref{eff_ham} and\,\eqref{eq:sc} exactly describe the dynamics of the interacting model in Eq.\,\eqref{ham_eq} at large $N$\,\cite{sotiriadis2010quantum}, so we are going to mostly refer to this limit in the following. However, our qualitative picture, including the existence of a novel form of prethermal universality, is expected to hold also at finite $N$. The effect of the quench dynamics on the effective potential is illustrated in Fig.~\ref{fig:quench}.

Following Refs.\,\cite{fring_solvable_2018, frith_time-dependence_2020} we reduce the non-Hermitian Hamiltonian\,\eqref{eff_ham} to a Hermitian one via the Dyson map $\eta(t)$, a time-dependent linear operator that maps energy eigenstates from the non-Hermitian system to eigenstates of a suitable Hermitian system. 
Expectation values are then evaluated with regard to a time-dependent metric which also preserves the norm of states.
In the case of the effective Hamiltonian in Eq.\,\eqref{eff_ham} one can follow a procedure analogous to the one developed in Ref.\,\cite{frith_time-dependence_2020} to derive the  Dyson map $\eta(t)$ and obtain the Hermitian Hamiltonian $h_\mathrm{eff}(t) = \eta(t) H_\mathrm{eff}(t) \eta^{-1}(t) + \ir \qty[\del_t \eta(t)] \eta^{-1}(t)$ describing the dynamics of our system\,(see Appendix). This Dyson map can then be expressed in terms of four time-dependent parameters $\theta_{\vk}, \varphi_{\vk}, \xi_{\vk}, \zeta_{\vk}$ that must be solutions to a system of coupled nonlinear differential equations such that the non-Hermitian part of $h_\mathrm{eff}(t)$ vanishes for all times $t>0$. Since these equations are also coupled to the self-consistency equation \eqref{eq:sc}, it is generally impossible to solve them exactly.

However, an explicit solution can be obtained in the prethermal regime of the dynamics, which occurs at small times $\tau = \sqrt{\abs{\lambda}} t \ll 1$. There, the differential equations for the Dyson map parameters decouple from the self-consistency equation, and the Hermitian Hamiltonian in momentum space $k$ (ignoring terms of $\order{\tau^3}$ and higher) simplifies to
\begin{align}
  h_\mathrm{eff}(t) = \half \sum_{i=1,2} \int_k 
  \qty[\Pi_{i,\vk} \Pi_{i,-\vk} + 
  \omega_{i,k}(t)^2 \phi_{i,\vk} \phi_{i,-\vk}] \ ,
\end{align}
where
\begin{align}
  \label{eq:eff_freq}
  \omega_{i,k}(t)^2 = r_{i,\text{eff}}(t) - \lambda^2 t^2 + k^2 .
\end{align}
We thus find that the two fields are effectively decoupled, and in the following we suppress the index $i$ and write $\phi = \phi_1$.  One can express the field in terms of creation and annihilation operators as $\phi_{\vk}(t) = f_{\vk}(t) a_{\vk} + f_{\vk}^*(t) a_{-\vk}^\dag$. The Heisenberg equation of motion under the action of $h_\mathrm{eff}(t)$ then takes the form
\begin{align}\label{eq:mode_eom}
  \ddot{f}_{\vk} + \omega_k(t)^2 f_{\vk} = 0
\end{align}
for the mode functions $f_{\vk}(t) = f_{-\vk}(t)$ with initial conditions $f_{\vk}(0) = 1/\sqrt{2\omega_k(0)}$ and $\dot{f}_{\vk}(0) = -\ir\sqrt{\omega_k(0)/2}$.
The self-consistency equation reads
\begin{align}\label{eq:sc_int}
  r_\mathrm{eff}(t) = \rc{} + \frac{u}{6} \int^\Lambda 
  \frac{\dd^d k}{(2\pi)^d} \abs{f_{\vk}(t)}^2 + \order{\tau^4} \ ,
\end{align}
where $\Lambda$ is a large-momentum cutoff introduced to avoid UV divergences.
The solution of the equation of motion \eqref{eq:mode_eom} subject to the constraint \eqref{eq:sc_int} provides the exact short-time dynamics.

\section{Scaling ansatz} At the critical point where the time-independent part of the effective mass vanishes (see below), the effective mass has to follow a scaling ansatz in analogy to Ref.\,\cite{maraga_aging_2015},
\begin{align}\label{eq:mass_scaling}
\tilde{r}_\mathrm{eff}(t) = \frac{a}{t^2} \qq{for} \Lambda t \gg 1 
\end{align}
due to the absence of any finite length scale at the critical point. Here, $\tilde{r}_\mathrm{eff}(t) = r_\mathrm{eff}(t) - \lambda^2 t^2$ is the effective mass including the effects of the imaginary coupling, and the coefficient $a$ is dimensionless. Introducing the scaling ansatz \eqref{eq:mass_scaling} into the equation of motion \eqref{eq:mode_eom} we obtain the analytical solution $f_{\vk}(t) = \sqrt{kt}[A_{\vk} J_\alpha(kt) + B_{\vk} J_{-\alpha}(kt)]$, where $\alpha = \sqrt{1/4 - a}$\,\cite{maraga_aging_2015}. While the coefficients $A_{\vk}, B_{\vk}$ depend in general on the nonuniversal functional form of $r_\mathrm{eff}(t)$ at short times, one can find scaling forms for them in the case of a deep quench. In that case, the condition $\Omega_{01} \gg \Lambda$ holds and the initial conditions for the mode functions become independent of $k$ at leading order. At times $t \simeq \Lambda^{-1}$, one can then use asymptotic forms for the Bessel functions at small arguments to find the scaling behavior $A_{\vk} = A(k/\Lambda)^{-1/2+\alpha}$ and $B_{\vk} = B(k/\Lambda)^{-1/2-\alpha}$, where $A$ and $B$ are two complex coefficients.

Let us now determine the value of the quantity $\alpha$ (and therefore also $a$). This is done by solving the self-consistency equation \eqref{eq:sc_int}. It can be shown that the modulus of the mode function $\abs{f_{\vk}}^2$ acquires the scaling form
\begin{align}\label{eq:mode_scaling}
  \abs{f_{\vk}(t)}^2 \simeq \abs{A}^2 
  \qty(\frac{\Lambda}{k})^{1 + 2\alpha} kt\, J_\alpha(kt)^2 \ ,
\end{align}
since the term proportional to $\abs{A}^2$ dominates the dynamics both in the short and long time limits\,\cite{maraga_aging_2015}. Combining equations \eqref{eq:sc_int}, \eqref{eq:mass_scaling}, and \eqref{eq:mode_scaling} yields
\begin{align}
  \label{eq:cond}
  \frac{a}{t^2} + \lambda^2 t^2 = \rc{} + 
  \frac{\Omega_d}{(2\pi)^d} \frac{u}{6} \abs{A}^2 \Lambda^d 
  R_{d,\alpha}(\Lambda t) \ ,
\end{align}
where $\Omega_d = 2\pi^{d/2} / \Gamma(d/2)$ denotes the $d$-dimensional solid angle, and we have defined the function $R_{d,\alpha}(x) = x\int_0^1 \dd y\, y^{d-1-2\alpha} J_{\alpha}(xy)^2$.
Expanding $R_{d,\alpha}$ for $x \gg 1$ we find
\begin{align}
  \begin{split}
    R_{d,\alpha}(x) &= W_{d,\alpha}(x) + c^{(0)}_{d,\alpha}\ + \\
    &\qquad\frac{c^{(1)}_{d,\alpha}}{x^{d-1-2\alpha}} + 
    \frac{c^{(2)}_{d,\alpha}}{x^2} + \order{\inv{x^4}} \ ,
  \end{split}
\end{align}
where $W_{d,\alpha}$ is a fast oscillating function due to the sharp momentum cutoff in the integral and is thus nonuniversal. The value of the bare mass $\rc{}$ at the critical point has to be determined by requiring a vanishing final mass, so that the constant contribution proportional to $c^{(0)}_{d,\alpha}$ in Eq.\,\eqref{eq:cond} is cancelled by $\rc{}$. On the other hand, matching the $t^2$ term on the l.h.s.\ of Eq.\,\eqref{eq:cond} requires $1 + 2\alpha - d = 2$, so that the term proportional to $c^{(1)}_{d,\alpha}$ also becomes quadratic in time.
In that case $c^{(1)}_{d,\alpha} = 4[\pi d(d+2)]^{-1}$, which leads to the constraint
\begin{align}\label{eq:lambda}
  \frac{\lambda^2}{u} = \frac{4\abs{A}^2 \Lambda^{d+2}}{
  3(2\sqrt{\pi})^d \pi d(d+2)\Gamma(d/2)} \ .
\end{align}

The interpretation of the constraint \eqref{eq:lambda} is rather straightforward: the open nature of the system leads to a quadratic $\sim t^{2}$ term in the effective mass $\tilde{r}_\mathrm{eff}(t)$, see Eq.\,\eqref{eq:eff_freq}, which is proportional to the imaginary coupling $\sim \lambda^{2}$ and has to be compensated by the coherent evolution caused by the interaction term $\sim u$. As a result, universal scaling can only be observed in the dark states, where the effect of non-Hermiticity is temporarily removed. As time grows, higher-order terms in $R_{d,\alpha}$ become relevant and non-Hermitian dynamics resumes, leading to the effectively prethermal nature of the dark state.

Thus, at variance with the isolated case\,\cite{maraga_aging_2015}, universal scaling in the non-Hermitian system can only be observed at fixed interaction strengths according to Eq.\,\eqref{eq:lambda}. When this constraint is fulfilled, the resulting values read $\alpha = (d+1)/2$ and $a = -(d^2 + 2d)/4$. At short times $t \ll k^{-1}$, the time evolution of correlations at criticality is characterized by the critical exponent $\theta$, sometimes referred to as the \emph{initial-slip} exponent, which can be defined via the short-time scaling of the retarded Green's function
\begin{align}\label{eq:gr_def}
  \delta_{\vk,-\vk'} \ir G_\mathrm{R}(k, t, t') &= \Theta(t-t')
		\ev{\comm{\phi_{\vk}(t)}{\phi_{\vk'}(t')}} \\
	&\propto -t \qty(\frac{t'}{t})^\theta \qq{for} t' \ll t \ll k^{-1} \ .
	\nonumber
\end{align}
Evaluating $G_\mathrm{R}(k,t,t')$ explicitly at short times $t'<t\ll k^{-1}$ yields the scaling form (see Ref.\,\cite{maraga_aging_2015})
\begin{align}\label{eq:gr_explicit}
	G_\mathrm{R}(k,t,t') \simeq 
	-\frac{t}{2\alpha} \qty(\frac{t'}{t})^{1/2 - \alpha} 
		\qty[1 - \qty(\frac{t'}{t})^{2\alpha}] \ .
\end{align}
Comparing Eqs.\,\eqref{eq:gr_def} and \eqref{eq:gr_explicit}, it follows that prethermal correlations scale with the exponent $\theta = -d/2$. Interestingly, and at variance with the Hermitian case, the scaling ansatz in Eq.\,\eqref{eq:mass_scaling} does not apply when the condition in Eq.\,\eqref{eq:lambda} on $\lambda$ and $u$ is not fulfilled. 
%%%%%%%%%%%%%%%%%%%%%%%%%%%%%%%%%%
\begin{figure}
\includegraphics[width=\linewidth]{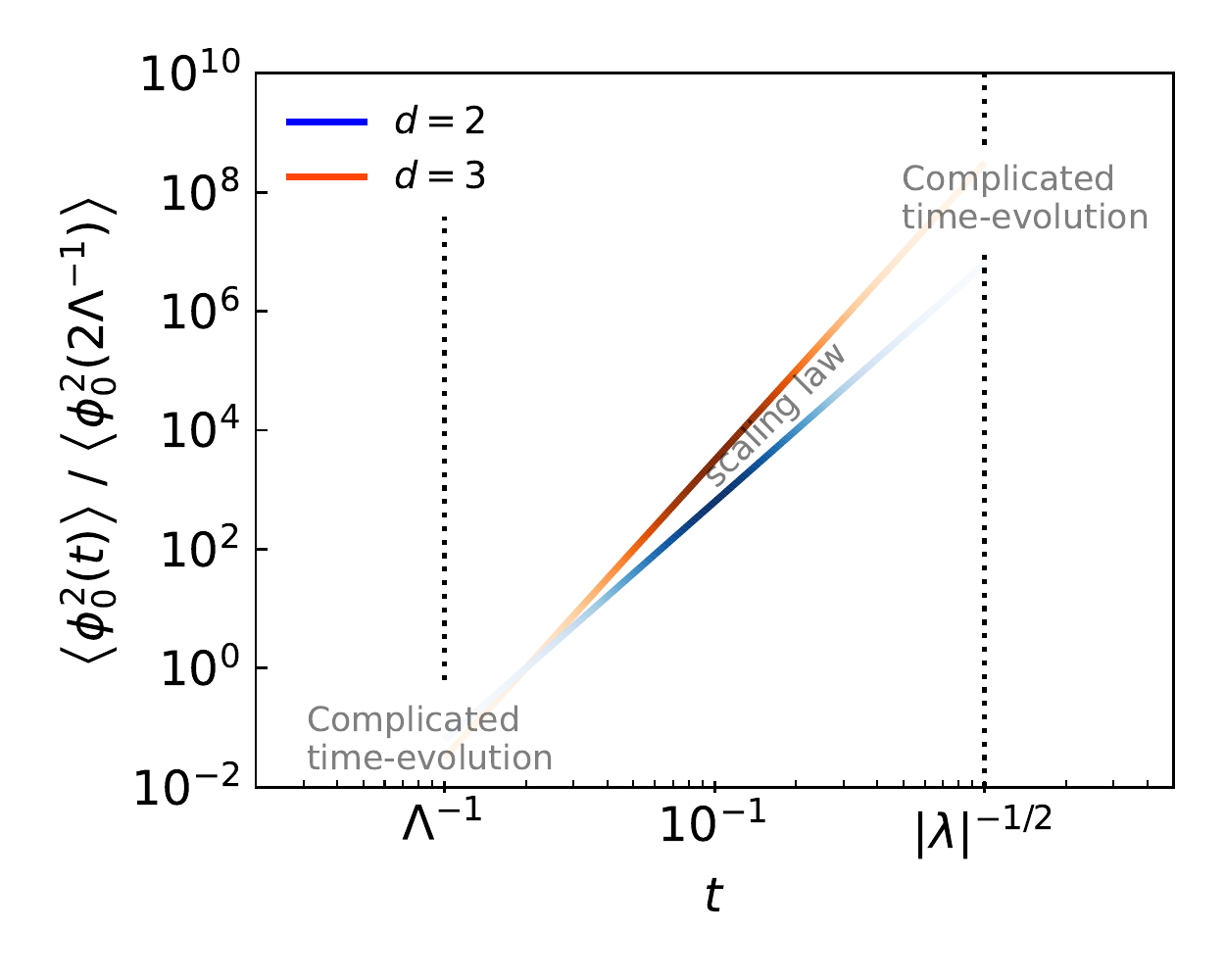}
\caption{Time evolution of the zero-momentum component $\ev*{\phi_0^2(t)} = \eval*{\ev*{\phi_{\vk}(t) \phi_{-\vk}(t)}}_{k=0}$ (corresponding to the magnetization) in $d=2$ and $d=3$ dimensions as obtained by Eq.\,\eqref{eq:mode_scaling}, shown for $\Lambda = 100$ and $\lambda = 1$. In both cases, correlations grow algebraically in the regime $\Lambda^{-1} \ll t \ll \abs{\lambda}^{-1/2}$. Outside this regime the correlations have a more complicated time dependence and cannot be computed analytically.}
\label{fig:time}
\end{figure}

\section{Conclusion}
In conclusion, we introduced a non-Hermitian model of \PT-symmetrically coupled $O(N)$ fields and showed---using the Dyson map formalism---that it is equivalent to an effective Hermitian model of two decoupled fields. The resulting differential equations were solved in the limit $t \ll \abs{\lambda}^{-1/2}$ and we demonstrated that the effective Hermitian Hamiltonian acquires a mass term quadratically decreasing with time. Due to the decoupling, the dynamics is governed by a single field whose bare mass parameter is quenched towards the critical value, such that the long-time limit (time-independent part) of the effective mass vanishes.

\begin{figure}
\includegraphics[width=\linewidth]{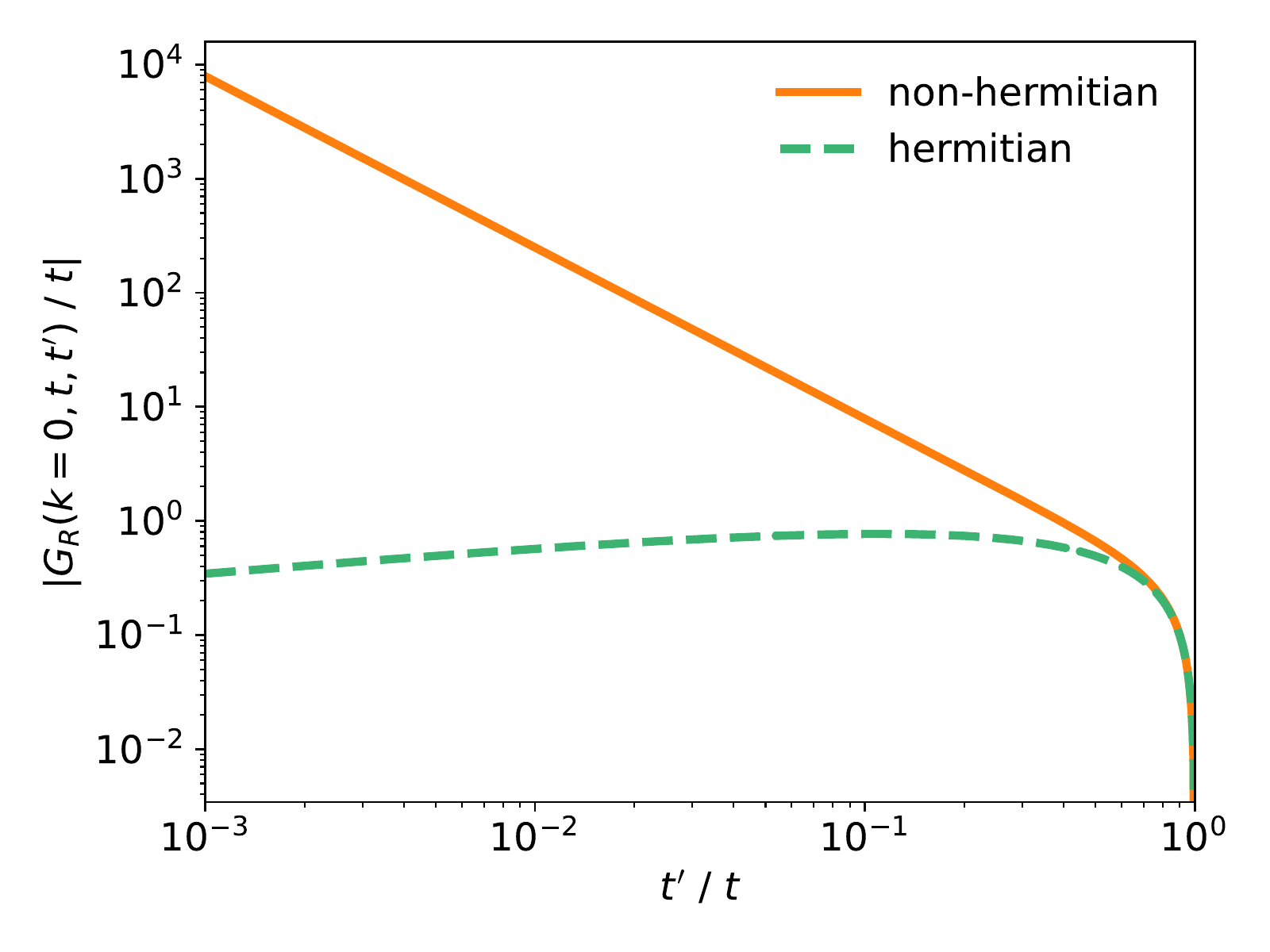}
\caption{A comparison between the analytical scaling form Eq.\,\eqref{eq:gr_explicit} of the retarded Green's functions of our non-Hermitian model and the Hermitian ($\lambda=0$) model in $d=3$ dimensions. In that case, $\alpha=2, \theta=-3/2$ for the non-Hermitian model while $\alpha=1/4, \theta=1/4$ for the Hermitian model.}
\label{fig:gr}
\end{figure}

According to dimensional analysis, the existence of the critical point is
directly connected with the $\tilde{r}_{\rm eff}\sim t^{-2}$ scaling of the time-dependent effective mass, making it necessary to fulfill the condition between non-Hermiticity and interaction in Eq.\,\eqref{eq:lambda}. The imaginary $\lambda$ term in the Hamiltonian \eqref{non_herm_term} produces an additional $t^{2}$ factor, which in general prevents the universal prethermal scaling observed in isolated systems\,\cite{maraga_aging_2015}. However, the interaction $u$ may be tuned to compensate the non-Hermitian contribution to the mass scaling: this restores universal short-time dynamics for $\Lambda^{-1} \ll t \ll \abs{\lambda}^{-1/2}$. Consequently, the slip exponent is different from the isolated case, and we find $\theta=-d/2$ as illustrated in Fig.~\ref{fig:time} and Fig.~\ref{fig:gr}.

Unsurprisingly, the smaller the imaginary coupling $\lambda$, the larger the time window where universal short-time dynamics is observed. Accordingly, this dynamical universality occurs entirely within the \PT-symmetric region of the phase diagram, well away from the exceptional points occurring at $|\lambda|>|r_{\rm eff}|$\,(see Appendix and \cite{syed_equilibrium_inpreparation}).  An interesting question for future studies concerns the impact of the exceptional points on the dynamics.

 It is worth noting that the Hermitian result found in Ref.\,\cite{maraga_aging_2015} is not obtained in the $\lambda\to 0$ limit of Eq.\,\eqref{eq:gr_explicit}. Indeed, the self-consistency condition in Eq.\,\eqref{eq:lambda} imposes a correlated limit where the ratio $\lambda^{2}/u$ is kept constant. In this context, the self-consistency condition in Eq.\,\eqref{eq:lambda} requires that criticality is reached by tuning two independent parameters $u$ and $\lambda$, as would be necessary to reach a  conventional equilibrium tricritical point. Thus, this scaling phenomenon can be regarded a third order prethermal fixed point and its observation should not pose further challenges than in other forms of tricriticality\,\cite{mortazavi1991nonequilibrium}.

The current analysis reveals the possibility to observe novel scaling phenomena arising from energy exchange of the quantum system with its environment. Such scaling phenomena can only occur if many-body correlations compensate for the changes in the time evolution produced by the non-Hermitian nature of the Hamiltonian. As a result, the system enters a long-lived dark state where dissipation effects are suppressed, while prethermal scaling is modified due to non-Hermiticity. Since these results are universal for $O(N)$ models, we expect these prethermal dark states to occur in a wide range of physical systems. Thus, the present analysis yields the fundamental theoretical insight to guide experimental realisations of prethermal scaling beyond the setting of isolated quantum systems described so far\,\cite{erne_universal_2018,eigen_universal_2018,prufer_observation_2018}.

\emph{Acknowledgements:} N.\,D.\ acknowledges fruitful exchange with A. Chiocchetta. This work is supported by the Deutsche Forschungsgemeinschaft (DFG, German Research Foundation), project-ID 273811115 (SFB1225 ISOQUANT) and under Germany’s Excellence Strategy EXC2181/1-390900948 (the Heidelberg STRUCTURES Excellence Cluster).

\bibliographystyle{apsrev4-2}
\bibliography{references}

\appendix*
\section{Derivation of the Dyson map}
We start with the Fourier transformed effective Hamiltonian
\begin{widetext}
\begin{align}
	H_\mathrm{eff}(t) = \int\! \frac{\dd^d k}{(2\pi)^d} 
	\Biggl\{ \half \sum_{i=1,2} 
		\biggl[\Pi_{i,\vk} \Pi_{i,-\vk}\ + 
			\qty(\reff_i(t) + k^2) \phi_{i,\vk} \phi_{i,-\vk}
		\biggr] +
			\ir\lambda \phi_{1,\vk} \phi_{2,-\vk} \Biggr\} \ ,
\end{align}
\end{widetext}
which can be written in vector form as
\begin{align}
	H_\mathrm{eff}(t) = \half \int_k 
		\bigl[\vb{P}_{\vk} \vdot \vb{P}_{-\vk} + 
		\vb*{\Phi}_{\vk} \vdot \Vv_{\vk}(t) \vb*{\Phi}_{-\vk}\bigr]
\end{align}
using $\vb{P}_{\vk} = (\Pi_{1,\vk}, \Pi_{2,\vk})$ and $\vb*{\Phi}_{\vk} = (\phi_{1, \vk}, \phi_{2,\vk})$ as well as the matrix
\begin{align}
	\Vv_{\vk}(t) = \mqty(\reff_1(t) + k^2 & \ir \lambda \\ 
		\ir \lambda & \reff_2(t) + k^2) \ .
\end{align}

\subsection{Time-independent case}
In the case where $\reff_i(t) = \reff_i$ is constant, one can find the eigenvalues of $\Vv_{\vk}$ which are
\begin{align}
	\omega_{\pm,k}^2 = \reff_+ + k^2 \pm  
		\sqrt{\qty(\reff_-)^2 - \lambda^2} \ ,
\end{align}
where $\reff_{\pm} = \qty(\reff_1 \pm \reff_2)/2$.
Evidently, the spectrum is real when the condition
\begin{align}
	\abs{\reff_-} > \abs{\lambda}
\end{align}
is fulfilled.
In other words, there are exceptional points at $\lambda = \lambda_\mathrm{e}^\pm$ with
\begin{align}
	\lambda_\mathrm{e}^\pm = \pm \reff_- \ ,
\end{align}
where \PT-symmetry is spontaneously broken.

\subsection{Time-dependent case}
We now go back to our time-dependent quenched system where the effective mass is a function of time.
To treat our problem we use the so-called time-dependent Dyson map $\eta(t)$ which is an operator mapping eigenstates $\ket{\Psi(t)}$ of $H_\mathrm{eff}(t)$ to eigenstates $\ket{\psi(t)} = \eta(t) \ket{\Psi(t)}$ of a Hermitian Hamiltonian $h_\mathrm{eff}(t)$\,\cite{frith_time-dependence_2020, fring_solvable_2018}.
Expectation values $\ev{\cdots} = \ev{\cdots \rho}{\Psi(t)}$ have to be evaluated with regard to the metric $\rho = \eta^\dag \eta$.
This procedure works regardless of $\lambda$, i.e., both in the \PT-symmetric and in the \PT-broken regime.
To ensure $h_\mathrm{eff}(t)$ is really the generator of time translations in the Hermitian system fulfilling the Schrödinger equation $\ir \del_t \ket{\psi(t)} = h_\mathrm{eff}(t)\ket{\psi(t)}$, the time-dependent Dyson equation
\begin{align}\label{eq:tdde}
	h_\mathrm{eff}(t) = \eta(t) H_\mathrm{eff}(t) \eta^{-1}(t) + 
		\ir \qty[\del_t \eta(t)] \eta^{-1}(t)
\end{align}
has to hold for $\eta(t)$ and $h_\mathrm{eff}(t)$.
The common way to solve this equation is to introduce an ansatz for $\eta(t)$, compute the right-hand side of \eqref{eq:tdde}, set $h_\mathrm{eff}(t) = h_\mathrm{eff}^\dag(t)$, and solve the resulting differential equations.

\subsubsection{Dyson map ansatz}
We make an ansatz similar to the one of Ref.\,\cite{frith_time-dependence_2020}. 
In order to do that we introduce the operators
\begin{subequations}
\begin{align}
	L_{\vk} &= \phi_{1,\vk} \Pi_{2,-\vk} - \phi_{2,\vk} \Pi_{1,-\vk} \\
	K_{\vk} &= \phi_{1,\vk} \Pi_{2,-\vk} + \phi_{2,\vk} \Pi_{1,-\vk} \\
	X_{\vk} &= \phi_{1,\vk} \phi_{2,-\vk} \\
	Z_{\vk} &= \Pi_{1,\vk} \Pi_{2,-\vk} 
\end{align}
\end{subequations}
whose commutation relations with the fields are
\begin{widetext}
\begin{subequations}
\begin{align}
	\comm{L_{\vk}}{\phi_{1,\vk'}} &= \ir \phi_{2,\vk} \delta_{\vk,\vk'} ,& 
	\comm{L_{\vk}}{\Pi_{1,\vk'}} &= \ir \Pi_{2,-\vk} \delta_{\vk,-\vk'} ,\\
	\comm{L_{\vk}}{\phi_{2,\vk'}} &= -\ir \phi_{1,\vk} \delta_{\vk,\vk'} ,&
	\comm{L_{\vk}}{\Pi_{2,\vk'}} &= -\ir \Pi_{1,-\vk} \delta_{\vk,-\vk'} ,\\
	\comm{K_{\vk}}{\phi_{1,\vk'}} &= -\ir \phi_{2,\vk} \delta_{\vk,\vk'} ,& 
	\comm{K_{\vk}}{\Pi_{1,\vk'}} &= \ir \Pi_{2,-\vk} \delta_{\vk,-\vk'} ,\\
	\comm{K_{\vk}}{\phi_{2,\vk'}} &= -\ir \phi_{1,\vk} \delta_{\vk,\vk'} ,&
	\comm{K_{\vk}}{\Pi_{2,\vk'}} &= \ir \Pi_{1,-\vk} \delta_{\vk,-\vk'} ,\\
	\comm{Z_{\vk}}{\phi_{1,\vk'}} &= -\ir \Pi_{2,-\vk} \delta_{\vk,-\vk'} ,&
	\comm{X_{\vk}}{\Pi_{1,\vk'}} &= \ir \phi_{2,-\vk} \delta_{\vk,-\vk'} ,\\
	\comm{Z_{\vk}}{\phi_{2,\vk'}} &= -\ir \Pi_{1,\vk} \delta_{\vk,\vk'} ,&
	\comm{X_{\vk}}{\Pi_{2,\vk'}} &= \ir \phi_{1,\vk} \delta_{\vk,\vk'} 
\end{align}
\end{subequations}
\end{widetext}
while all other commutators with single fields vanish,
and the contracted operators
\begin{subequations}
\begin{align}
	L(t) &= \int_k \theta_{\vk}(t) L_{\vk} \\
	K(t) &= \int_k \varphi_{\vk}(t) K_{\vk} \\
	X(t) &= \int_k \xi_{\vk}(t) X_{\vk} \\
	Z(t) &= \int_k \zeta_{\vk}(t) Z_{\vk} 
\end{align}
\end{subequations}
whose commutation relations with the fields are
\begin{subequations}
\begin{align}
	\comm{L}{\phi_{1,\vk}} &= \ir \thetak \phi_{2,\vk}  ,& 
	\comm{L}{\Pi_{1,\vk}} &= \ir \thetak \Pi_{2,-\vk}  ,\\
	\comm{\phi_{2,\vk}}{L} &= \ir \thetak \phi_{1,\vk} ,&
	\comm{\Pi_{2,\vk}}{L} &= \ir \thetak \Pi_{1,-\vk}  ,\\
	\comm{K}{\phi_{1,\vk}} &= -\ir \phik \phi_{2,\vk} ,& 
	\comm{K}{\Pi_{1,\vk}} &= \ir \phik \Pi_{2,-\vk}  ,\\
	\comm{K}{\phi_{2,\vk}} &= -\ir \phik \phi_{1,\vk} ,&
	\comm{K}{\Pi_{2,\vk}} &= \ir \phik \Pi_{1,-\vk}  ,\\
	\comm{Z}{\phi_{1,\vk}} &= -\ir \zetak \Pi_{2,-\vk}  ,&
	\comm{X}{\Pi_{1,\vk}} &= \ir \xik \phi_{2,-\vk}  ,\\
	\comm{Z}{\phi_{2,\vk}} &= -\ir \zetak \Pi_{1,\vk} ,&
	\comm{X}{\Pi_{2,\vk}} &= \ir \xik \phi_{1,\vk}  \ .
\end{align}
\end{subequations}
while all other commutators vanish again.
The parameters $\thetak, \phik , \xik, \zetak$ are functions of time with the property $\thetak = \theta_{-\vk}, \phik = \varphi_{-\vk}, \xik = \xi_{-\vk}, \zetak = \zeta_{-\vk}$.
The commutation relations for $L$ imply
\begin{subequations}\label{eq:angmom}
\begin{align}
	\er^L \vb*{\Phi}_{\vk} \er^{-L} &= \mqty(\cosh\thetak & 
		\ir\sinh\thetak \\ -\ir\sinh\thetak & 
		\cosh\thetak) \vb*{\Phi}_{\vk} \\
	\er^L \vb{P}_{\vk} \er^{-L} &= \mqty(\cosh\thetak & 
		\ir\sinh\thetak \\ -\ir\sinh\thetak & 
		\cosh\thetak) \vb{P}_{\vk} \ .
\end{align}
\end{subequations}
From the commutation relations for $K$ it also follows that
\begin{subequations}
\begin{align}
	\er^K \vb*{\Phi}_{\vk} \er^{-K} &= \mqty(\cos\phik & -\ir\sin\phik \\
		-\ir\sin\phik & \cos\phik) \vb*{\Phi}_{\vk} \\
	\er^K \vb{P}_{\vk} \er^{-K} &= \mqty(\cos\phik & \ir\sin\phik \\
		\ir\sin\phik & \cos\phik) \vb{P}_{\vk} \ .
\end{align}
\end{subequations}
Finally, we have
\begin{align}
	\comm{K_{\vk}}{L_{\vk'}} = 2\ir C_{\vk} \delta_{\vk,\vk'}
\end{align}
with
\begin{align}
	C_{\vk} = \phi_{1,\vk} \Pi_{1,-\vk} - \phi_{2,\vk} \Pi_{2,-\vk} 
\end{align}
and
\begin{align}
	\comm{K_{\vk}}{C_{\vk'}} = 2\ir L_{\vk} \delta_{\vk,\vk'} \ .
\end{align}

Let $\eta$ then be of the form
\begin{align}
	\eta(t) = \er^{Z(t)} \er^{K(t)} \er^{X(t)} \er^{L(t)} \ .
\end{align}
First, we calculate the derivative term
\begin{widetext}
\begin{align}
	\ir \qty[\del_t \eta(t)] \eta^{-1}(t) = 
	\ir \int_k \qty{\dot{\zeta}_{\vk} Z_{\vk} + 
		\dot{\varphi}_{\vk} \er^Z K_{\vk} \er^{-Z} + 
		\dot{\xi}_{\vk} \er^Z \er^K X_{\vk} \er^{-K} \er^{-Z} + 
		\dot{\theta}_{\vk} \er^Z \er^K \er^X L_{\vk} 
			\er^{-X} \er^{-K} \er^{-Z}} \ ,
\end{align}
starting with the first nontrivial subterms
\begin{subequations}\label{eq:deriv1}
\begin{align}
	\er^Z K_{\vk} \er^{-Z} &= 
	K_{\vk} - \ir \zeta_{\vk} \qty(\Pi^1_{\vk} \Pi^1_{-\vk} + 
		\Pi^2_{\vk}\Pi^2_{-\vk}) =
	K_{\vk} - \ir \zeta_{\vk} \vb{P}_{\vk} \vdot \vb{P}_{-\vk} \\
	\er^K X_{\vk} \er^{-K} &= \qty(\phi_{1,\vk} \cos\varphi_{\vk} - 
		\ir \phi_{2,\vk} \sin\varphi_{\vk})
		\qty(\phi_{2,-\vk} \cos\varphi_{\vk} - 
		\ir \phi_{1,-\vk} \sin\varphi_{\vk}) \nonumber \\ &=
	X_{\vk} \cos^2 \varphi_{\vk} - X_{-\vk} \sin^2 \varphi_{\vk} - 
	\half[\ir] \vb*{\Phi}_{\vk} \vdot \vb*{\Phi}_{-\vk} \sin 2\varphi_{\vk} \\ 
	\er^{X} L_{\vk} \er^{-X} &=
	L_{\vk} + \ir \xi_{\vk} \qty(\phi_{1,\vk} \phi_{1,-\vk} - 
		\phi_{2,\vk} \phi_{2,-\vk}) =
	L_{\vk} + 
	\ir \xi_{\vk}	\vb*{\Phi}_{\vk} \vdot \sigma_z \vb*{\Phi}_{-\vk} \ ,
\end{align}
\end{subequations}
where $\sigma_z = \operatorname{diag}(1,-1)$ denotes the $z$ Pauli matrix. 
In the next step we encounter the terms
\begin{subequations}\label{eq:deriv2}
\begin{align}
	\er^Z X_{\vk} \er^{-Z} &= 
	X_{\vk} - 
		\ir \zeta_{\vk} 
			\qty(\Pi_{2,\vk} \phi_{2,-\vk} + \phi_{1,\vk} \Pi_{1,-\vk}) - 
		\zeta_{\vk}^2 Z_{-\vk} \nonumber\\ &= 
	X_{\vk} -
		\ir \zeta_{\vk} 
			\qty(\Pi_{2,\vk} \phi_{2,-\vk} + \phi_{1,\vk} \Pi_{1,-\vk}) - 
		\zeta_{\vk}^2 Z_{-\vk} \nonumber\\ &=
	X_{\vk} - 
		\ir \zeta_{\vk} D_{\vk} - \zeta_{\vk}^2 Z_{-\vk} - \zeta_{\vk}
	\\
	\er^Z \vb*{\Phi}_{\vk} \vdot \vb*{\Phi}_{-\vk} \er^{-Z} &= 
	\vb*{\Phi}_{\vk} \vdot \vb*{\Phi}_{-\vk} -
		\ir \zeta_{\vk} 
			\qty(\Pi_{2,\vk} \phi_{1,-\vk} + \phi_{1,\vk} \Pi_{2,-\vk} + 
		\Pi_{1,\vk} \phi_{2,-\vk} + \phi_{2,\vk} \Pi_{1,-\vk}) -
		\zeta_{\vk}^2 \vb{P}_{\vk} \vdot \vb{P}_{-\vk} \nonumber\\ &=
	\vb*{\Phi}_{\vk} \vdot \vb*{\Phi}_{-\vk} - 
		\ir\zeta_{\vk} \qty(K_{\vk} + K_{-\vk}) -
		\zeta_{\vk}^2 \vb{P}_{\vk} \vdot \vb{P}_{-\vk} 
	\\
	\er^K L_{\vk} \er^{-K} &= 
	L_{\vk} \cos 2\varphi_{\vk} + \ir C_{\vk} \sin 2\varphi_{\vk} 
	\\
	\er^K \vb*{\Phi}_{\vk} \vdot \sigma_z \vb*{\Phi}_{-\vk} \er^{-K} &=
	\vb*{\Phi}_{\vk} \vdot \sigma_z \vb*{\Phi}_{-\vk} \ ,
\end{align}
\end{subequations}
where $D_{\vk} = \phi_{1,\vk} \Pi_{1,-{\vk}} + \phi_{2,\vk} \Pi_{2,-{\vk}}$.
Lastly, we have
\begin{subequations}\label{eq:deriv3}
\begin{align}
	\er^Z L_{\vk} \er^{-Z} &= L_{\vk} + 
		\ir \zeta_{\vk} \vb{P}_{\vk} \vdot \sigma_z \vb{P}_{-\vk} \\
	\er^Z C_{\vk} \er^{-Z} &= C_{\vk} +
		\ir\zeta_{\vk}\qty(Z_{\vk} - Z_{-\vk}) \\
	\er^Z \vb*{\Phi}_{\vk} \vdot \sigma_z \vb*{\Phi}_{-\vk} \er^{-Z} &= 
	\vb*{\Phi}_{\vk} \vdot \sigma_z \vb*{\Phi}_{-\vk} - 
		\ir \zeta_{\vk} \qty(L_{\vk} + L_{-\vk}) + 
		\zeta_{\vk}^2 \vb{P}_{\vk} \vdot \sigma_z \vb{P}_{-\vk} \ .
\end{align}
\end{subequations}
\end{widetext}
Combining all of the equations above and taking only the non-Hermitian part of the derivative term yields
\begin{align}
\begin{split}
	\ir \qty[\del_t \eta(t)] \eta^{-1}(t) - \mathrm{h.c.} = 
	2\ir\int_k \bigl[
		\dot{\zeta}_{\vk} Z_{\vk} + \dot{\varphi}_{\vk} K_{\vk}\ + \\
		\dot{\xi}_{\vk} X_{\vk} \cos 2\varphi_{\vk} - 
		\zeta_{\vk}^2\dot{\xi}_{\vk} Z_{\vk} \cos 2\varphi_{\vk}\ - \\ 
		\zeta_{\vk} \dot{\xi}_{\vk} K_{\vk} \sin 2\varphi_{\vk} + 
		\dot{\theta}_{\vk} L_{\vk} \cos 2\varphi_{\vk} + 
		2\zeta_{\vk} \xi_{\vk} \dot{\theta}_{\vk} L_{\vk}
	\bigr] \ .
\end{split}
\end{align}

Next we look at the kinetic term $H_\mathrm{kin} = \half\int_k P_{\vk} \vdot P_{-\vk}$, which is left invariant by adjoint action of $\er^L$ due to the fact that the matrix in Eq.\,\eqref{eq:angmom} is orthogonal, and thus leaves scalar products invariant.
Adjointly applying $\er^X$ yields
\begin{widetext}
\begin{align}
	\er^X \vb{P}_{\vk} \vdot \vb{P}_{-\vk} \er^{-X} &=
	\vb{P}_{\vk} \vdot \vb{P}_{-\vk} + 
		\ir \xi_{\vk} \qty(\phi_{2,\vk} \Pi_{1,-\vk} + 
		\Pi_{1,\vk} \phi_{2,-\vk} + \phi_{1,\vk} \Pi_{2,-\vk} + 
		\Pi_{2,\vk} \phi_{1,-\vk}) - 
		\xi_{\vk}^2 \vb*{\Phi}_{\vk} \vdot \vb*{\Phi}_{-\vk}
	\nonumber \\ &=
	\vb{P}_{\vk} \vdot \vb{P}_{-\vk} + 
		\ir \xi_{\vk} \qty(K_{\vk} + K_{-\vk}) - 
		\xi_{\vk}^2 \vb*{\Phi}_{\vk} \vdot \vb*{\Phi}_{-\vk} \ .
\end{align}
Doing the same with $\er^K$, this transforms into
\begin{align}
	\vb{P}_{\vk} \vdot \vb{P}_{-\vk} \cos 2\varphi_{\vk} + 
	\ir \qty(Z_{\vk} + Z_{-\vk}) \sin 2\varphi_{\vk} - 
	\xi_{\vk}^2 \vb*{\Phi}_{\vk} \vdot 
		\vb*{\Phi}_{-\vk} \cos 2\varphi_{\vk}\ + \qquad \nonumber\\
	\ir \xi_{\vk}^2 \qty(X_{\vk} + X_{-\vk}) \sin 2\varphi_{\vk} + 
	\ir \xi_{\vk} \qty(K_{\vk} + K_{-\vk}) \ .
\end{align}
Since we will ultimately integrate over all $\vk$, we can use the symmetry of the coefficient functions and write
\begin{align}
	\vb{P}_{\vk} \vdot \vb{P}_{-\vk} \cos 2\varphi_{\vk} + 
	2\ir Z_{\vk} \sin 2\varphi_{\vk} -
	\xi_{\vk}^2 \vb*{\Phi}_{\vk} \vdot 
		\vb*{\Phi}_{-\vk} \cos 2\varphi_{\vk} + 
	2\ir \xi_{\vk}^2 X_{\vk} \sin 2\varphi_{\vk} +
	2\ir \xi_{\vk} K_{\vk} 
\end{align}
instead.
For the adjoint action of $\er^Z$ on these operators, we already know the transformation rules from Eqs.\,\eqref{eq:deriv1}, \eqref{eq:deriv2}, and \eqref{eq:deriv3}.
In total, the contribution of the kinetic term is therefore
\begin{align}
	\eta(t) H_\mathrm{kin} \eta^{-1}(t) - \mathrm{h.c.} = 2\ir \int_k \qty[
		\qty(1 - \zeta_{\vk}^2 \xi_{\vk}^2) Z_{\vk} \sin 2\varphi_{\vk}\ + 
		\xi_{\vk} 
			\qty(\zeta_{\vk} \xi_{\vk} \cos 2\varphi_{\vk} + 1) K_{\vk} + 
		\xi_{\vk}^2 X_{\vk} \sin 2\varphi_{\vk} ] \ .
\end{align}

Lastly, we transform the potential term $H_\mathrm{pot}(t) = \half \int_k \vb*{\Phi}_{\vk} \vdot \Vv_{\vk}(t) \vb*{\Phi}_{-\vk}$. 
It is left invariant by $\er^X$.
The two rotations via $\er^L$ and $\er^K$ yield
\begin{align}
	\eta(t) H_\mathrm{pot}(t) \eta^{-1}(t) = \half
	\er^Z \int_k \qty(
		v_{+,\vk} \vb*{\Phi}_{\vk} \vdot \vb*{\Phi}_{-\vk} + 
		v_{-,\vk} \vb*{\Phi}_{\vk} \vdot \sigma_z \vb*{\Phi}_{-\vk} + 
		2\ir w_{\vk} X_{\vk}) \er^{-Z} 
\end{align}
with
\begin{subequations}
\begin{align}
	v_{+,\vk} &= \qty(\reff_+ + k^2) \cos 2\varphi_{\vk} +
		\qty(2\lambda \cosh 2\theta_{\vk} + \reff_- \sinh 2\theta_{\vk}) 
		\sin 2\varphi_{\vk} \\
	v_{-,\vk} &= 2\lambda \sinh 2\theta_{\vk} + 
		\reff_- \cosh 2\theta_{\vk}\\
	w_{\vk} &= \qty(2 \lambda \cosh 2\theta_{\vk} + 
		\reff_- \sinh 2\theta_{\vk}) \cos 2\varphi_{\vk} -
		\qty(\reff_+ + k^2) \sin 2\varphi_{\vk} \ . 
\end{align}
\end{subequations}
Applying $\er^Z$ and taking only the non-Hermitian part yields
\begin{align}
	\eta(t) H_\mathrm{pot}(t) \eta^{-1}(t) - \mathrm{h.c.} = 2\ir\int_k
	\bigl[
		w_{\vk} X_{\vk} - \zeta_{\vk}^2 w_{\vk} Z_{\vk} -
		\zeta_{\vk} v_{+,\vk} K_{\vk} - \zeta_{\vk} v_{-,\vk} L_{\vk}
	\bigr] \ .
\end{align}
Because the operators involved in all the final expressions are linearly independent, we can compare coefficients to obtain the differential equations
\begin{subequations}
\begin{align}
	\dot{\zeta}_{\vk} - \zeta_{\vk}^2 \dot{\xi}_{\vk} \cos 2\varphi_{\vk} 
	&= 
		\zeta_{\vk}^2 w_{\vk} + 
		\qty(\zeta_{\vk}^2 \xi_{\vk}^2 - 1) \sin 2\varphi_{\vk} 
	\\
	\dot{\xi}_{\vk} \cos 2\varphi_{\vk} &= 
		-\xi_{\vk}^2 \sin 2\varphi_{\vk} - w_{\vk} 
	\\
	\dot{\varphi}_{\vk} - \zeta_{\vk} \dot{\xi}_{\vk} \sin 2\varphi_{\vk} 
	&=
		-\qty(1 + \zeta_{\vk} \xi_{\vk} \cos 2\varphi_{\vk}) \xi_{\vk} + 
		\zeta_{\vk} v_{+,\vk} 
	\\
	\qty(\cos 2\varphi_{\vk} + 2\zeta_{\vk} \xi_{\vk}) \dot{\theta}_{\vk} 
	&= 
		\zeta_{\vk} v_{-,\vk} \ .
\end{align}
\end{subequations}
Simplifying a bit, we get
\begin{subequations}\label{eq:dyson_eom}
\begin{align}
	\dot{\zeta}_{\vk} &= 
		-\sin 2\varphi_{\vk} \\
	\dot{\varphi}_{\vk} &= 
		\zeta_{\vk} \qty(\reff_+ + k^2 - \xi_{\vk}^2)\sec 2\varphi_{\vk} - 
		\xi_{\vk} \\
	\dot{\xi}_{\vk} &= 
		\qty(\reff_+ + k^2 - \xi_{\vk}^2) \tan 2\varphi_{\vk} -
		\reff_- \sinh 2\theta_{\vk} - 2\lambda \cosh 2\theta_{\vk} \\
	\dot{\theta}_{\vk} &= 
		\frac{\qty(2\lambda\sinh2\theta_{\vk} + 
		\reff_- \cosh2\theta_{\vk})\zeta_{\vk}}{\cos2\varphi_{\vk} + 
		2\zeta_{\vk} \xi_{\vk}} \ .
\end{align}
\end{subequations}
\end{widetext}

\subsubsection{Hermitian Hamiltonian}
Remarkably, the resulting Hermitian Hamiltonian decouples into two independent parts
\begin{align}
	h_\mathrm{eff}(t) = 
	\half \qty[h_\mathrm{eff}(t) + h_\mathrm{eff}^\dag(t)] = 
	h_{1,+}(t) + h_{2,-}(t)
\end{align}
where
\begin{widetext}
\begin{align}
	h_{i,\pm}(t) = \int_{k} 
		\qty{\inv{2M_{\pm,\vk}(t)} \Pi_{i,\vk} \Pi_{i,-\vk} + 
		\half M_{\pm,\vk}(t) \omega_{\pm,\vk}^2(t) 
			\phi_{i,\vk} \phi_{i,-\vk} \mp
		\half g_{\vk}(t) 
			\qty[\phi_{i,\vk} \Pi_{i,-\vk} + \Pi_{i,\vk} \phi_{i,-\vk}]} \ .
\end{align}
The coefficients are
\begin{subequations}\label{eq:params}
\begin{align}
	M_{\pm,\vk}^{-1} &= \cos 2\varphi_{\vk} + 
		\zeta_{\vk}^2 \Gamma_{\pm,\vk} \\
	\omega_{\pm,\vk}^2 &= \Gamma_{\mp,\vk} M_{\pm,\vk}^{-1} \\
	g_{\vk} &= \dot{\theta}_{\vk} \sin 2\varphi_{\vk} \ ,
\end{align}
\end{subequations}
where
\begin{align}\label{eq:gamma}
	\Gamma_{\pm,\vk} = \qty(\reff_+ + k^2 - \xi_{\vk}^2) \sec 2\varphi_{\vk} \mp 
		\frac{2\lambda\sinh2\theta_{\vk} + 
			\reff_- \cosh2\theta_{\vk}}{\cos2\varphi_{\vk} + 
				2\zeta_{\vk} \xi_{\vk}} \cos 2\varphi_{\vk} \ . 
\end{align}
\end{widetext}

\subsubsection{Approximate solution to the differential equations}
Solving Eq.\,\eqref{eq:dyson_eom} analytically is generally not possible because of the coupling to the self-consistency equation through $\reff_\pm$.
Hence, we will only provide an approximate solution to \eqref{eq:dyson_eom} for short times $t \ll \abs{\lambda}^{-1/2}$.
To do this, we introduce the parameter $\tau = \abs{\lambda}^{1/2}t$ and expand the functions $\theta_{\vk}, \varphi_{\vk}, \xi_{\vk}$, and $\zeta_{\vk}$ as power series in $\tau$.

Inserting $t = \abs{\lambda}^{-1/2} \tau$ into Eq.\,\eqref{eq:dyson_eom} yields
\begin{widetext}
\begin{subequations}\label{eq:tau_eom}
\begin{align}
	\zeta_{\vk}' &= 
		-\frac{\sin 2\varphi_{\vk}}{\abs{\lambda}^{1/2}} \\
	\varphi_{\vk}' &= \frac{
		\zeta_{\vk} \qty(\reff_+ + k^2 - \xi_{\vk}^2)\sec 2\varphi_{\vk} - 
		\xi_{\vk}}{\abs{\lambda}^{1/2}} \\
	\xi_{\vk}' &= \frac{
		\qty(\reff_+ + k^2 - \xi_{\vk}^2) \tan 2\varphi_{\vk} -
		\reff_- \sinh 2\theta_{\vk}}{\abs{\lambda}^{1/2}} -
		\frac{2\lambda}{\abs{\lambda}^{1/2}} \cosh 2\theta_{\vk} \\
	\theta_{\vk}' &= 
		\frac{\qty(2\lambda\sinh2\theta_{\vk} + 
		\reff_- \cosh2\theta_{\vk})\zeta_{\vk}}{
		\abs{\lambda}^{1/2} \qty(\cos2\varphi_{\vk} + 
		2\zeta_{\vk} \xi_{\vk})} \ ,
\end{align}
\end{subequations}
\end{widetext}
where $(\cdots)' = \dv{\tau} (\cdots)$.
Our initial condition is $\theta_{\vk}(0) = \varphi_{\vk}(0) = \xi_{\vk}(0) = \zeta_{\vk}(0) = 0$, because we started in a Hermitian ground state.
From Eq.\,\eqref{eq:tau_eom} it is then easy to see that all first derivatives vanish at $\tau=0$ except for $\xi_{\vk}'(0) = -2\lambda / \abs{\lambda}^{1/2}$.
At second order, the only nonvanishing term is $\varphi_{\vk}''(0) = -\xi_{\vk}'(0) / \abs{\lambda}^{1/2} = 2\lambda / \abs{\lambda}$.
We thus obtain the solution
\begin{subequations}\label{eq:tau_sol}
\begin{align}
	\zeta_{\vk}(t) &= \order{\tau^3} \\
	\varphi_{\vk}(\tau) &= \frac{\lambda}{\abs{\lambda}} \tau^2 +
		\order{\tau^3} \\
	\xi_{\vk}(\tau) &= -\frac{2\lambda}{\abs{\lambda}^{1/2}} \tau + 
		\order{\tau^3} \\
	\theta_{\vk}(\tau) &= \order{\tau^3} \ .
\end{align}
\end{subequations}
Inserting this into Eqs.\,\eqref{eq:gamma} and \eqref{eq:params} yields $\Gamma_{\pm,\vk}(\tau) = \reff_+ \mp \reff_- + k^2 - \frac{\lambda^2}{\abs{\lambda}}\tau^2 + \order{\tau^3}$ and
\begin{subequations}
\begin{align}
	M_{\pm,\vk}^{-1}(\tau) &= 1 + \order{\tau^4} \\
	\omega_{\pm,\vk}(\tau)^2 &= \reff_+ \pm \reff_- + k^2 -
		\frac{\lambda^2}{\abs{\lambda}}\tau^2 + \order{\tau^3} \\
	g_{\vk}(\tau) &= \order{\tau^4} \ .
\end{align}
\end{subequations}
This procedure thus results in the final Hamiltonian
\begin{align}
	h_\mathrm{eff}(t) = \half \sum_{i=1,2} \int_k 
		\qty[\Pi_{i,\vk} \Pi_{i,-\vk} + 
		\omega_{i,k}(t)^2 \phi_{i,\vk} \phi_{i,-\vk}] \ ,
\end{align}
where $\omega_{i,k}(t)^2 = \omega_{\pm,\vk}(t)^2 = \reff_i(t) + k^2 - \lambda^2 t^2$.

\subsubsection{Self-consistency equation}
Let us denote the Schrödinger representation ground state of $h_\mathrm{eff}(t)$ as $\ket{\psi_0(t)}$, and the one of $H_\mathrm{eff}(t)$ as $\ket{\Psi_0(t)}$.
The expectation value inside the self-consistency equation is
\begin{align}
	\ev{\phi_{i,\vk} \phi_{i,-\vk}} &= 
	\ev{\qty(\phi_{i,\vk})^\dag \phi_{i,\vk} \rho}{\Psi_0(t)} \nonumber\\ &=
	\ev{\eta^{-1}\qty(\phi_{i,\vk})^\dag \phi_{i,\vk} \eta}{\psi_0(t)} 
		\nonumber\\ &=
	\ev{\qty(\phi_{i,\vk}\eta^{-1})^\dag \eta \eta^{-1} 
		\phi_{i,\vk} \eta}{\psi_0(t)} \nonumber\\ &=
	\ev{\qty(\eta \phi_{i,\vk} \eta^{-1})^\dag 
		\eta^{-1} \phi_{i,\vk} \eta}{\psi_0(t)} \ ,
\end{align}
where we used $\ket{\Psi_0(t)} = \eta^{-1} \ket{\psi_0(t)}$, and that $\eta^\dag = \eta$ in our case.
Using the previous approximation, we obtain $\eta^{-1} \phi_{i,\vk} \eta = \phi_{i,\vk} + \ir \lambda t^2 \sum_{j=1,2} (1-\delta_{ij}) \phi_{j,\vk} + \order{\tau^3}$, and therefore
\begin{widetext}
\begin{align}
	\ev{\phi_{i,\vk} \phi_{i,-\vk}} = 
	\ev{\phi_{i,\vk} \phi_{i,-\vk}}{\psi_0(t)} + 
	\ir\lambda t^2 \sum_{j=1,2} (1- \delta_{ij}) \ev{\phi_{j,\vk} \phi_{i,-\vk} + 
		\phi_{i,\vk} \phi_{j,-\vk}}{\psi_0(t)} + 
	\order{\tau^4} \ .
\end{align}
The term quadratic in time vanishes after taking the expectation value and we are left with the expression from the main part.
\end{widetext}

\end{document}